# MUSIC GENRE CLASSIFICATION USING SPECTRAL ANALYSIS AND SPARSE REPRESENTATION OF THE SIGNALS

Mehdi Banitalebi-Dehkordi　Amin Banitalebi-Dehkordi

## ABSTRACT

In this paper, we proposed a robust music genre classification method based on a sparse FFT based feature extraction method which extracted with discriminating power of spectral analysis of non-stationary audio signals, and the capability of sparse representation based classifiers. Feature extraction method combines two sets of features namely short-term features (extracted from windowed signals) and long-term features (extracted from combination of extracted short-time features). Experimental results demonstrate that the proposed feature extraction method leads to a sparse representation of audio signals. As a result, a significant reduction in the dimensionality of the signals is achieved. The extracted features are then fed into a sparse representation based classifier (SRC). Our experimental results on the GTZAN database demonstrate that the proposed method outperforms the other state of the art SRC approaches. Moreover, the computational efficiency of the proposed method is better than that of the other Compressive Sampling (CS)-based classifiers.

*Index Terms*— Feature Extraction, Compressive Sampling, Genre Classification

## 1. INTRODUCTION

Audio classification provides useful information for understanding the content of both audio and audio-visual recordings. Audio information can be classified from different points of view. Among them, the generic classes of music have attracted a lot of attention [1]. Low bit-rate audio coding is an application that can benefit from distinguishing music classes [2]. In the previous studies, various classification schemes and feature extraction methods have been used for this purpose. Most of the music genre classification algorithms resort to the so-called bag-of-features approach [3-4], which models the audio signals by their long-term statistical distribution of short-time features. Features commonly exploited for music genre classification can be roughly classified into timbral texture, rhythmic, pitch content ones, or their combinations [3-4]. Having extracted descriptive features, pattern recognition algorithms are employed for their classification into genres. These features can be categorized into two types, namely short-term and long-term features. The short-term features are derived from a short segment (such as a frame). In contrast, long-time features usually characterize the variation of spectral shape or beat information within a long segment. The short and long segments are also referred to as "analysis window" and "texture window" respectively [5]. In short-time estimates, the signal is partitioned into successive frames using small sized windows. If an appropriate window size is chosen the signal within each frame can be considered as a stationary signal. The windowed signal is then transformed into another representation space in order to achieve good discrimination and/or energy compaction properties. The length of a short term analysis window depends on the signal type. In music classification, the length of a short-term window is influenced by the adopted audio coding scheme [6]. A music data window of length less than 50 ms is usually referred to as a short-term window [7].

Non-stationary signals such as audio signals can be modeled as the product of a narrow bandwidth low-pass process modulating a higher bandwidth carrier [7]. The low-pass content of these signals cannot be effectively captured by using a too short analysis window. To improve the deficiency of the short-term feature analysis, we propose a simple but very effective method for combining the short-term and long-term features. Optimum selection of the number of features also plays an important role in this context. Too few features can fail to encapsulate sufficient information while too many features usually degrade the performance since they may be irrelevant. Moreover, too many features could also entail excessive computation downgrading the system's efficiency. As a result, we need an effective method for feature extraction and/or selection. Compressive or Sparse Sampling turns out to be an appropriate tool for such a purpose. In this paper, a Compressive Sampling (CS) based classifier which is based on the theory of sparse representation and reconstruction of signals is presented for music genre classification. The innovative aspect of our approach lies in the adopted method of combining short-term (extracted from windowed signals) and long-term (extracted from combination of extracted short-time features) signal characteristics to make a decision.

An important issue in audio classification algorithms which has not been widely investigated is the effect of background noise on the classification performance. In fact, a classification algorithm trained using clean sequences may fail to work properly when the actual testing sequences contain background noise with a certain level of signal-to-noise ratio (SNR). For practical applications wherein environmental sounds are involved in audio classification tasks, noise robustness is an essential characteristic of the processing system. We show that the proposed feature extraction and data classification algorithm is robust to the background noise.

The rest of this paper is organized as follows. In the next section, the proposed feature extraction algorithm is described. The corresponding CS-based classifier is then introduced in Section 3. The experimental settings and results are detailed in Section 4 leading to conclusions in Section 5.

## 2. COMPRESSIVE SENSING BACKGROUND

We sample the speech signal x(t) at the Nyquist rate and process it in frames of N samples. Each frame is then a $N \times 1$ vector x, which can be represented as $x = \Psi X$, where $\Psi$ is a $N \times N$ matrix whose columns are the similarly sampled basis functions $\psi_i(t)$, and $X$ is the vector that chooses the linear combinations of the basis functions. $X$ can be thought of as x in the domain of $\Psi$, and it is $X$ that is required to be sparse for compressed sensing to perform well. We say that $X$ is K-sparse if it contains only $K$ non-zero elements. In other words, x can be exactly represented by the linear combination of $K$ basis functions.

It is also important to note that compressed sensing will also recover signals that are not truly sparse, as long these signals are highly compressible, meaning that most of the energy of x is contained in a small number of the elements of $X$.

Given a measurement sample $Y \in R^M$ and a dictionary $D \in R^{M \times N}$ (the columns of $D$ are referred to as the atoms), we seek a vector solution $X$ satisfying:

$$\min \|X\|_0 \quad s.t. \quad Y = DX \quad (1)$$

In above equation $\| \|_0$ (known as $l_0$ norm), is the number of non-zero coefficient of $X$ [5].

In Fig. 1, consider a signal $X$ with Length $N$ that has $K$ non-zero coefficients on sparse basis matrix $\Psi$, and consider also an $M \times N$ measurement basis matrix $\Phi$, $M \ll N$ where the rows of $\Phi$ are incoherent with the columns of $\Psi$. In matrix notation, we have $X = \Psi \theta$, in which $\theta$ can be approximated using only $K \ll N$ non-zero entries. The CS theory states that such a signal $X$ can be reconstructed by taking linear, non-adaptive measurement as follows [10]:

$$Y = \Phi X = \Phi \Psi \theta = A \theta \quad (2)$$

where $Y$ represents an $M \times 1$ sampled vector, $A = \Phi \Psi$ is an $M \times N$ matrix. The reconstruction is equivalent to finding the signal's sparse coefficient vectors $\theta$, which can be cast into a $l_0$ optimization problem:

$$\min \|\theta\|_0 \quad s.t. \quad y = \Phi \Psi \theta = A \theta \quad (3)$$

Generally, (3) is NP-hard, and this $l_0$ optimization is replaced with an $l_1$ optimization [10] as follows:

$$\min \|\theta\|_1 \quad s.t. \quad Y = \Phi \Psi \theta = A \theta \quad (4)$$

From (4) we will recover $\theta$ with high probability if enough measurements are taken. In general, the $l_p$ norm is defined as:

$$\|a\|_p = (\sum_j |a_j|^p)^{\frac{1}{p}}. \quad (5)$$

Equation (4) can be reformulated as:

$$\min_\theta \|Y - \Phi \Psi \theta\|_2 \quad s.t. \quad \|\theta\|_0 = K, \quad (6)$$

There are a variety of algorithms to perform the reconstructions in (4) and (6); in this paper we make use of OMP [11] to solve (6).

OMP is a relatively-efficient iterative algorithm that produces one component of $\hat{\theta}$ each iteration, and thus allows for simple control of the sparsity of $\hat{\theta}$. As the true sparsity is often unknown, the OMP algorithm is run for a pre-determined number of iterations, $K$, resulting in $\hat{\theta}$ being K-sparse.

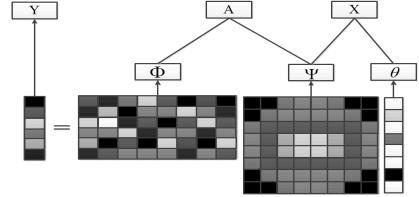

Figure 1. The measurement of Compressive Sampling

## 3. FEATURE EXTRACTION

In many of music genre classification algorithms authors used speech signal to determine music classics, this has effect on computational complexity. In [1-4], the computational complexity of algorithms is high, since it uses the high dimensional received signals. In [11], authors also used the high dimensional received signals, but they trying to reduce computational complexity by a feature extraction process that select the useful data for music genre classification.

Timbral texture features are frequently used in various music information retrieval systems [8]. Some timbral texture features which are widely used for audio classification have been summarized in Table I [1]. Among them, MFCC[1], spectral centroids, spectral roll off, spectral flux and zero crossings are short-time features, thus their statistics are computed over a texture window. The low-energy feature is a long-time feature.

In the previous speech processing researches, various feature extraction methods have been used. Scheirer and Slaney proposed features such as 4-Hz modulation energy and spectral centroids [2]. Various content-based features have been proposed for applications such as sound classification and music information retrieval (MIR) [3-4]. These features can be categorised into two types, namely short-term and long-term features. The short-term features are derived from a short segment (such as a frame). In contrast, long-time features usually characterise the variation of spectral shape or beat information within a long segment.

---
[1] Mell Frequency Coefficient

Non-stationary signals such as audio signals can be modelled as the product of a narrow bandwidth low-pass process modulating a higher bandwidth carrier [7]. The low-pass content of these signals cannot be effectively captured by using a too short analysis window. To improve the deficiency of the short-term feature analysis, we propose a simple but very effective method for combining the short-term and long-term features. Optimum selection of the number of features also plays an important role in this context. Too few features can fail to encapsulate sufficient information while too many features usually degrade the performance since they may be irrelevant. Moreover, too many features could also entail excessive computation downgrading the system's efficiency. As a result, we need an effective method for feature extraction and/or selection. Compressive or Sparse Sampling turns out to be an appropriate tool for such a purpose.

TABLE I. TIMBORAL TEXTURE FEATURE [11]

| Feature | Description |
|---|---|
| MFCC | Representation of the spectral characteristics based on Mel-frequency scaling |
| Spectral centroid | The centroid of amplitude spectrum |
| Spectral roll off | The frequency bin below which 85% of the spectral distribution is concentrated |
| Spectral flux | The squared difference of successive amplitude spectrums |
| Zero crossings | The number of time domain zero crossings of the music signal |
| Low-energy | The percentage of analysis windows that have energy less than the average energy across the texture window. |

In many audio classification systems, a spectrogram or other joint time-frequency representations are used for classifying audio signals. In these approaches, a feature vector is obtained by applying the Fast Fourier Transform (FFT) to the windowed audio signals. In our proposed method, the spectrogram is used as the starting point of the feature extraction process. In the first step, the audio signal is partitioned using a Hamming window function. In the next step, the FFT is applied to the audio signal segments for extracting short-term features. Then, the amplitudes of the FFT coefficients of each frame are normalized by dividing them by their maximum value. Now we have short-term feature vectors. In the fourth step, a new feature vector is formed in which each component is the sum of the normalized coefficients of the corresponding frame. As already mentioned, the audio signals are not stationary over a relatively long time interval. However, in the constructed feature space, the statistical properties of the resulting data become almost constant and can be considered as stationary. In the next step, the FFT of the vector composed of the feature samples of this stationary data is computed to extract long-term feature. This second FFT provides a sparse representation. Applying an amplitude filter a sparse feature vector is obtained [9]. In step six, by random sampling, we obtained used samples for music genre classification. Number of used samples obtained from simulations.

In this feature extraction process, the number of features is reduced from the number of samples to the number of frames. Moreover, the feature vector is sparse. In the next section, we show how in our proposed classifier the compressive sampling concept is adopted in order to effectively reduce the number of features further.

The whole feature extraction procedure can be described as a six-stage process as shown in Figure 2.

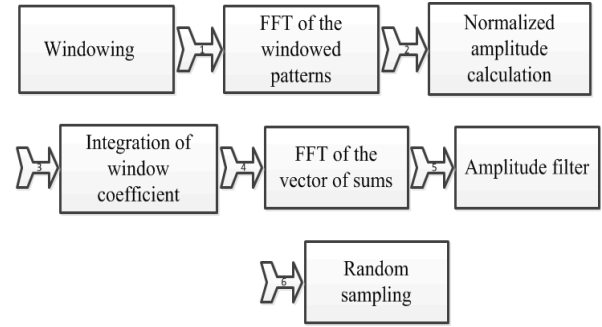

Figure 2. The proposed feature extraction algorithm

## 4. COMPRESSIVE SAMPLING BASED CLASSIFICATION

We consider the problem of music genre classification using a sparse representation based classifier in this part. For this purpose, we use a novel feature extraction method which combines the short-term features (output of part 2 in Figure (2)) and long-term features (output of part 5 in Figure (2)). The method leads to sparse representation of the audio signals. By compressive sampling and sparse signal representation theory, we present a CS-based classifier for music genre classification. So, we present a modified form of the Sparse Representation based Classifier [11]. It will show that the proposed method outperforms all state of the art approaches. Moreover, the computational efficiency of the method is better than that of the other Compressive Sampling (CS)-based classifiers.

In part 2 we said that the Compressed Sensing can be categorized as a type of dimensionality reduction technique. This theory states that for a given degree of residual error, ε, under some conditions, CS guarantees the success of recovering a given signal from a small number of samples [10, 13] which are acquired using a set of measurement vectors. The CS problem is expressed as follows: Given a measurement sample $y \in R^M$ and a dictionary $\varphi \in R^{M \times N}$ (the columns of φ are referred to as the atoms), we seek a vector solution x satisfying:

$$\min \; \|x\|_0 \quad s.t. \quad y = \varphi x \qquad (7)$$

Where $\|x\|_0$ is known as the $l_0$ norm i.e. the number of non-zero coefficients of $x$.

Unfortunately solving (7) is both numerically unstable and NP-complete. Therefore, the $l_1$ optimization is used to replace the above $l_0$ optimization:

$$\min \quad \|x\|_1 \quad s.t. \quad y = \varphi x \qquad (8)$$

Several iterative algorithms have been proposed to solve this minimisation problem (Greedy Algorithms such as Orthogonal Matching Pursuit (OMP) or Matching Pursuit (MP) and Non-convex local optimization like FOCUSS algorithm [11]. Based on this theory an important issue is that $x$ should be a sparse vector in which there are only a limited number of non-zero elements.

If it is not sparse in the original space it is assumed that the signal can be represented based on a few basis vectors. As in Figure 1, consider a signal $x$ (length $N$) that is $K$-sparse in the sparse basis matrix $\psi$, and consider also an $M \times N$ measurement basis matrix $\varphi$, where $M \ll N$ ($M$ is far less than $N$) and the rows of $\varphi$ are incoherent with the columns of $\psi$. In term of matrix notation, we have $x = \psi \theta$, in which $\theta$ can be approximated using only $K \ll N$ nonzero entries. The CS theory states that such a signal $x$ can be reconstructed by taking only $M = O(K \log N)$ linear, non-adaptive measurement as follows:

$$y = \varphi x = \varphi \psi \theta = A \theta \qquad (9)$$

Where y represents an $M \times 1$ sampled vector and $A = \varphi \psi$ is an $M \times N$ matrix. The reconstruction process is equivalent to finding the signal's sparse coefficient vectors $\theta$, which can be cast into the $l_0$ or $l_1$ optimisation problem.

Now, let the dimension of the extracted feature be denoted as $M$, and the extracted feature vector (feature extraction method illustrated in part 3) of the $j-th$ music in the $i-th$ class as $v_{i,j} \in R^M$. Moreover, let assume there are sufficient training samples for the $i-th$ class $\varphi_i = [v_{i,1}, v_{i,2}, ..., v_{i,n_i}] \in R^{M \times n_i}$. Then any new sample $y \in R^M$ from the same class can be written as below:

$$y = \sum_{i=1}^{n_i} \alpha_{i,n_i} v_{i,n_i} \qquad (10)$$

For some scalars $\alpha_{i,j} (j = 1, 2, ..., n_i)$. Since the membership $i$ (or the label) of the test sample is initially unknown, a new matrix $\varphi$ is defined for the entire training set as the concatenation of the $n$ training samples of all $K$ classes: $\varphi = [\varphi_1, \varphi_2, ..., \varphi_K]$. Then the linear representation of $y$ can be rewritten in terms of all training samples as:

$$y = \varphi x_0 \in R^M \qquad (11)$$

Where

$$x_0 \in [0, 0, ..., 0, \alpha_1, \alpha_2, ..., \alpha_n, 0, ..., 0]^T \in R^N \qquad (12)$$

Is coefficient vector whose entries are zero except those associated with the $i-th$ class and $N = \sum_{i=1}^{K} n_i$. As the non-zero entries of the vector $x_0$ encode the identity of the test sample, $y$, it is tempting to obtain it by solving Equation (7). Since $x_0$ is sparse this is called a sparse representation based classifier (SRC) [12].

Our proposed CS-based classifier is based on the principle of the SRC with an additional random measurement on the extracted features to reduce the number of dimensions of the signal which is fed into the classifier. The realisation of the algorithm is summarised as in Table 2.

TABLE II. THE PROPOSED CS-BASED CLASSIFIAR

|        | CS-based Classification |
|--------|-------------------------|
| Step 1 | Using the feature extraction process (Figure 2) calculate feature vector for training and test signals |
| Step 2 | Perform random measurement of the featurs using a Gaussian random matrix and normalise the data to have unit $l_2$ norm |
| Step 3 | Solve the l1 minimisation problem, eq. 4, for a test data |
| Step 4 | For each class, considering eq. 7 set the relevant entries of $x_0$ to zero and compute an approximation of the data using eq. 6 |
| Step 5 | Calculate the error for each class and select the class with the minimum error |

It should be noted that in the proposed algorithm by a random measurement of the feature vector, the computational complexity of the method is much lower than that of the basic SRC one. Moreover, the random measurements reduce the risk of over-fitting of the classifier to the training samples which define the transformation matrix. In other works in this field, many other non-linear dimensionality reduction methods are applied. The Local Coordinates Alignment (LCA) and Non-Negative Matrix Factorization (NMF) are two examples [9].

## 4. EXPERIMENTAL RESULTS

Our experimental study for music genre classification was performed on the GTZAN dataset which are widely used in this area [11, 13]. The GTZAN consists of the following ten genre classes: Classical, Country, Disco, Hip-Hop, Jazz, Rock, Blues, Reggae, Pop, and Metal. Each genre class contains 100 audio recordings of 30 seconds, with the sampling rate of 44.1 kHz and resolution of 16 bits. To evaluate the proposed, we set up all the experimental parameters to be as close as possible to those used in [8]. In particular, the recognition rate is obtained from 5-fold cross validation. As mentioned earlier, in the first step of our feature extraction method the music signal is partitioned using a Hamming window. A Hamming window of 100ms duration with 50 percent overlap was applied in our experiments.

First, we wanted to investigate the feasibility and robustness of the proposed feature extraction algorithm. Figure (3) shows plots of the extracted feature vectors related to two music signals from the same class. In this Figure, first row (sample 1 and sample 2), shows signal 1 and signal 2 from same class; second row (feature 1 and feature 2), illustrates output of part 3 of proposed feature extraction algorithm in Figure (2); and third row (feature 1 and feature 2), illustrates our proposed feature extraction method. From this Figure we can say that feature vectors extracted with our proposed feature extraction method for signals from same class, have similar feature vectors. Figure (4) contains similar plots of signals randomly chosen from different classes. In this Figure, first row (sample 1, sample 2 and sample 3), shows signal 1, signal 2 and signal 3 from different classes; second row (feature 1, feature 2 and feature 3), illustrates output of part 3 of proposed feature extraction algorithm in Figure (2); and third row (feature 1, feature 2 and feature 3), illustrates our proposed feature extraction method. From this Figure we can say that feature vectors extracted with our proposed feature extraction method for signals from different classes, have different feature vectors. This Figure shows that feature vectors extracted from different similar classes by our algorithm reveals class labels. These examples graphically demonstrate the within-class compactness and between-class separability properties of the extracted features.

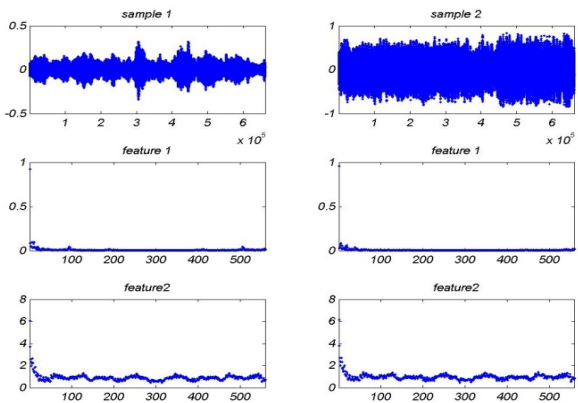

Figure 3. Within class variations of the extracted features

A few experiments have also been conducted in order to gauge the robustness of the proposed algorithm against additive noise. Figures (5) show the associated plots for features extracted from noisy signals respectively. As Figure (5) shows, with the noisy signal, although the SNR is very low, the extracted features are very similar to that of the clean signal. Also, as expected, since the phase information is not involved in the feature extraction process, there is no sensitivity to the phase shift. In this Figure, first row (sample 1 and sample 2), shows signal 1 (source signal) and signal 2 (source signal with Gaussian noise); second row (feature 1 and feature 2), illustrates output of part 3 of proposed feature extraction algorithm in Figure (2); and third row (feature 1 and feature 2), illustrates our proposed feature extraction method.

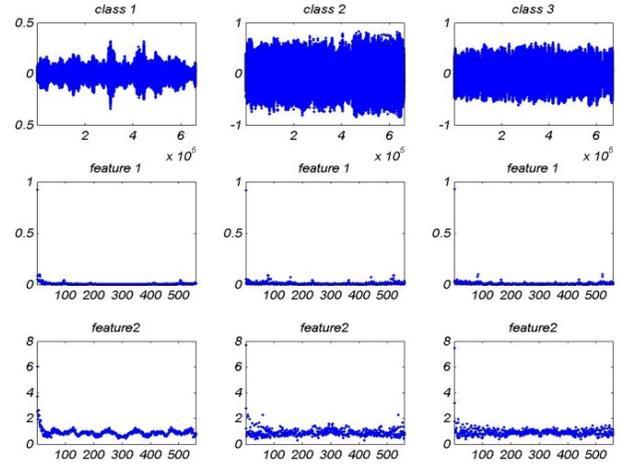

Figure 4. Between classes variations of the extracted features

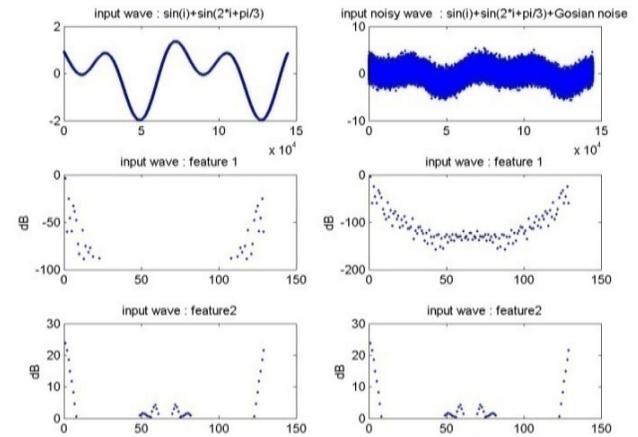

Figure 5. Robustness of the feature extraction method against additive noise

In the next step, the classification accuracy of the proposed system was evaluated and compared to a few state of the art SRC-based approaches. The adopted methods are the topology preserving non-negative matrix factorization (TPNMF), non-negative tensor factorization (NTF), multi linear Principal Component Analysis (MPCA) and General Tensor Discriminate Analysis (GTDA) [9]. Table III, contains the associated recognition rate results. It can be seen that, the proposed method outperforms all the SRC-based approaches. Moreover, the length of the feature vector of the proposed algorithm is considerably lower than that of the SRC-based approaches, demonstrating the effectiveness of the sparse sampling representation in extracting the features.

Table III addresses the problem of genre classification using Compressive Sampling. A CS recovery is applied on short-term (extracted from windowed signals) and long-term features (extracted from combination of extracted short-time features) that are related for genre classification. The measurement vectors are trained on labelled sets, then the classification is performed by computing the approximation of unknown samples with each class-specific features.

TABLE III. CLASSIFICATION ACCURACIES ACHIEVED BY VARIOUS METHODS ON GTZAN DATASET

| Method | Accuracy | Feature dimension |
|---|---|---|
| **Our Algorithm** | 95.7 | 35 |
| TPNMF+SRC | 91.4 | 125 |
| NTF+SRC | 90.3 | 125 |
| MPCA+SRC | 87.7 | 208 |
| GTDA+SRC | 90.1 | 208 |

Table IV shows the associated results when the number of training samples for each class varies between 1 and 99 samples (out of the total 100 samples). It can be seen that the error first rapidly decreases and it then remains almost at a same level when a large enough number of training samples become available. In this Table FFT illustrates output of part 3 of proposed feature extraction algorithm in Figure (2) and 2nd-fft illustrates output of proposed feature extraction algorithm.

TABLE IV. PERCENT OF ERROR IN CLASSIFICATION RESULTS FOR DIFFERENT NUMBER OF TEST DATA

| Test data number | 1 | 10 | 20 | 30 | 40 | 50 | 70 | 80 | 100 |
|---|---|---|---|---|---|---|---|---|---|
| Fft | 95 | 45.3 | 23.6 | 18.1 | 14.8 | 12.3 | 10.4 | 9.2 | 8.5 |
| 2nd-fft | 19.4 | 10.5 | 8.5 | 6.9 | 4.9 | 2.7 | 1.4 | 1 | 0.8 |

## 5. CONCLUSIONS

The problem of music genre classification using a sparse representation based classifier was considered. For this purpose, a novel feature extraction method which combines the short-term features (extracted from windowed signals) and long-term features (extracted from combination of extracted short-time features) was proposed. The method leads to sparse representation of the audio signals. Then, a modified form of the Sparse Representation based Classifier [11] was presented. It was shown that the proposed method outperforms all state of the art approaches. Moreover, the computational efficiency of the method is better than that of the other Compressive Sampling (CS)-based classifiers.